\documentclass[a4paper, oneside, twocolumn, notitlepage, 10pt]{extarticle_ecoc}
\usepackage{ecoc}
\usepackage{caption}
\usepackage{subcaption}
\usepackage{amsmath}
\usepackage{microtype}
\usepackage[style=ieee]{biblatex}
\begin{document}
\selectlanguage{english}    % Standard Language

%-------------------------------------------------- Title -----------------------------------------------------%

\title{On the Achievable Performance in the presence of Multiple Path Interference for Intra Data Center applications}%

%------------------------------------------------- Authors-----------------------------------------------------%

\author{
    Wing-Chau Ng\textsuperscript{(1)} and  Scott Yam\textsuperscript{(2)}
}

\maketitle                  % Create title and author

%------------------------------------------ Description of Authors ----------------------------------------------%

\begin{strip}
    \begin{author_descr}

        \textsuperscript{(1)} Carleton University
        \textsuperscript{(2)} Queens University
        \textcolor{blue}{\uline{scott.yam@queensu.ca}}

    \end{author_descr}
\end{strip}

% \setstretch{1.1}
%-------------------------------------------------- Footnote -------------------------------------------------------%
\renewcommand\footnotemark{}
\renewcommand\footnoterule{}
%\let\thefootnote\relax\footnotetext{text}

%-------------------------------------------------- Abstract ---------------------------------------------------------%

\begin{strip}
    \begin{ecoc_abstract}
       An accurate analytical form of the achievable bit error rate in the presence of multipath interference (MPI) is proposed for PAM4 for the first time, taking into account an ideal MPI estimate and compensation.  \textcopyright 2025 The Author(s)
    \end{ecoc_abstract}
\end{strip}

%-------------------------------------------------- Introduction Section -------------------------------------------------------%

\section{Introduction}
Four-level pulse amplitude modulation (PAM4) has been a promising technology ranging from wireline transceivers to short-reach fiber optics transmission systems, due to its low complexity, low cost, and power efficiency for AI infrastructure \cite{example:Marvell}.  Recent huge demands in data center connectivity for scale-up and scale-out have eventually driven the commercialization of 1.6 Tbps (200 Gbps/lane) PAM4 data center intraconnects (DCIs) by the end of 2025\cite{example:google}. In the next generation 400 Gbps per lane, the stringent device bandwidth will further shrink the margin of the intraconnect links for 2 km or above, while the optical amplitude-dependent impairments such as laser relative intensity noise (RIN), four-wave mixing (FWM) and multipath interference (MPI) may even fail the system requirements\cite{example:google}. Among all the optical impairments,  MPI is caused by the interferometric interaction between the signal and the reflected light upon photo-detection, whose dynamic is determined by laser phase noise\cite{example:stat}. There is hitherto no digital signal processing (DSP) algorithm to perfectly mitigate strong MPI in  digital receivers, and therefore DCI links need an extra margin to budget for the MPI penalty. 
\par 
Standard PAM4 DSP architecture consists of feedforward equalizer (FFE), decision feedback equalizer and maximum sequential likelihood detection \cite{example:Marvell}. The FFE is considered critical because the performance of the early-stage DSP block have to be stable and as optimal as possible to avoid "penalty magnification"  in the subsequent DSP stages.  A variety of research worked on how to cancel the MPI after the first FFE \cite{example:OFC2025}\cite{example:CLEO2023}.  Nevertheless, to achieve optimal performance, one has to jointly compensate ISI and MPI. It is because MPI dynamically displaces signal constellation, and therefore the conventional decision-directed LMS does not work optimally. This would lead to a system penalty that cannot be mitigated by any sorts of DSP algorithms. So far there has been no report to explore the best (i.e. achievable) BER performance that a single FFE can achieve.  
\par
In this contribution,  we start with formulating an MPI channel for a single-reflection case.  Then, we define the achievable performance and come up with an analytical expression of the achievable BER. Finally we present our numerical results, discuss about the insight on optimizing FFE performance and conclude this work. 

\section{Formulation}
Let us limit our scope to a single-reflection case. Upon photodetection with a unity photodiode responsivity (V/W) for simplicity, the photo-voltage of the MPI-impaired received optical signal \cite{example:stat} is
\begin{equation}\label{eq:model}
V_{pd}(t) ~= P_{s} + 2 \rho \sqrt{ P_{s}(t)P_{s}(t-\tau)} B(t) + n(t),   
\end{equation}
where $P_s$ is the optical signal power comprising zero-mean bipolar PAM4 data centered at an average power. Assume a perfectly-linear intensity modulator, the optical signal power is linearly proportional to the electrical data, assuming a unity conversion slope, i.e., $P_s(t) = V_{b} + V_d(t)$, where $V_{b}$ is the signal bias voltage offset from zero appeared after photodetection while $V_d = \{-3k, -k, k, 3k\}$ is the bipolar PAM4 data, where $k$ is the conversion from standard symbols $ \{-3, -1, 1, 3\}$ to electrical voltages.
\par
\begin{figure}[t!]
    \centering
    \includegraphics[height=3cm]{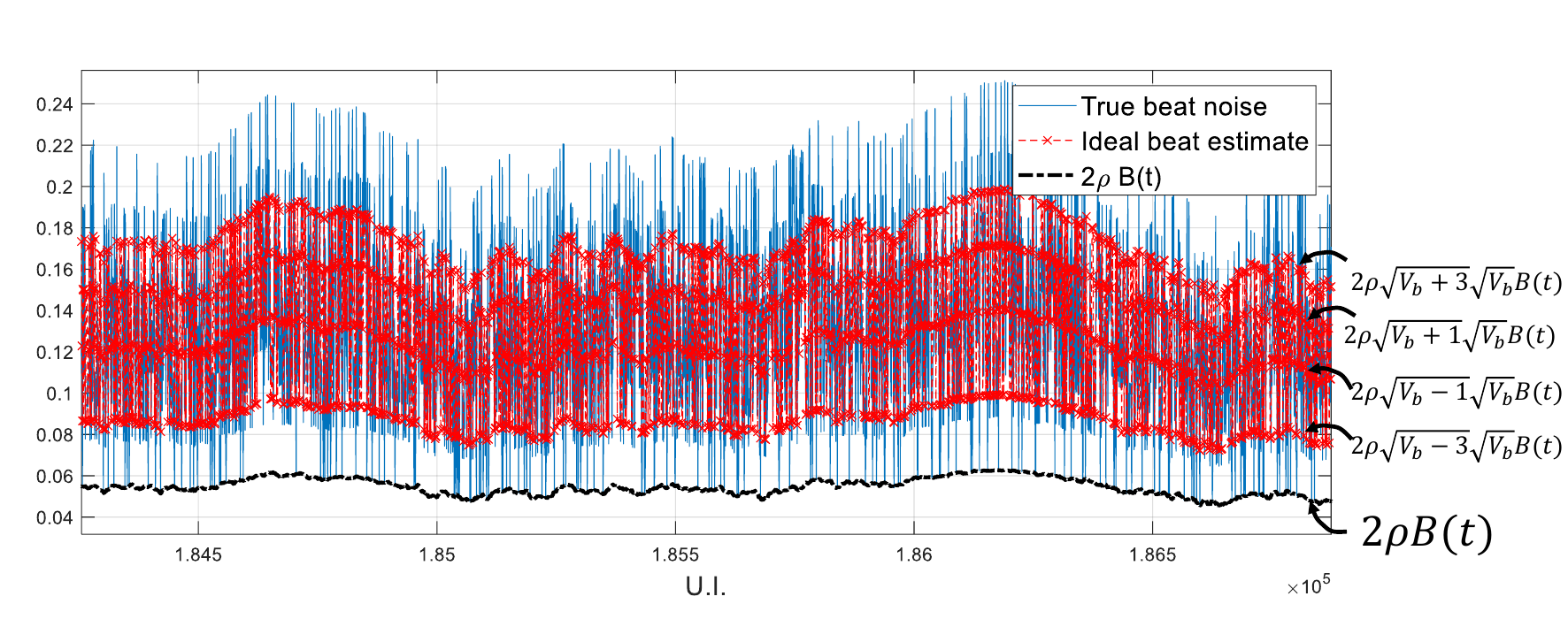}
    \caption{Bias fluctuation. Blue: True beat noise; Red: Best estimate of the true beat noise; Black: pure dynamic term}.
    \label{fig:dynamics}
\end{figure}

The second term in  Eq.~(\ref{eq:model}) is a beat term generated by mixing the signal, $P_s(t)$,  and the reflection, $P_s(t-\tau)$, delayed by $\tau$ and attenuated by the MPI (power) ratio, $\rho^2$, in the square-law device (photodetector), shown as blue in Fig. \ref{fig:dynamics}. This beat term causes signal DC fluctuation via the unknown delayed reflection as well as the phase-noise induced dynamics $B(t) = \cos{[\theta(t)-\theta(t-\tau)]}$ shown as dotted dashed in Fig. \ref{fig:dynamics} (bottom). The third term $n(t)$ is the additive white Gaussian noise (AWGN) to approximate the electrical or optical white noise. After mean removal and renormalization by $k$ at receiver (Rx) DSP,  the Rx signal becomes $y(t)= [V_{pd}(t)-V_b]/k$, which is compared with the transmitted symbol  $d(t) = V_d(t)/k  \in \{-3, -1, 1, 3\}$. Without loss of generality, set $k = 1$, the received signal can be written as 
\begin{equation}\label{eq:rxSignal}
\begin{split}
y(t) & = d(t) + 2 \rho \sqrt{ V_{b} + d(t)} \\  & \cdot \sqrt{V_{b} +d(t-\tau)} B(t) + n(t).
\end{split}
\end{equation}

\section{Proposed Achievable BER}
To define our "achievable" performance, firstly, one has to assume (i) for all time $t$, a perfect bipolar PAM4 decision $d(t) = \{-3, -1, 1, 3\}$ is available; (ii) there exists a perfect estimation method to extract the phase noise dynamics $B(t)$ in Eq.~(\ref{eq:rxSignal}). In practice, the delayed data $d(t-\tau)$ is unknown (unknown reflection path length), and is therefore uncorrelated with $d(t)$. This delayed reflection acts as a high frequency interference at data rate, that no ASIC-rate DSP algorithms can track accurately.  Given a certain MPI ratio, $\rho^2$, assume the existence of perfect prior information (i), (ii) and $V_b$, the ideal beat estimate can be written as 
\begin{equation}\label{eq:biasEst}
 \hat{b} =  2 \rho \sqrt{ V_{b} + d(t)} \cdot \sqrt{V_{b}} B(t), 
\end{equation}
 which is actually the mean of the bias fluctuation of each PAM4 level, represented by the four red curves in Fig. \ref{fig:dynamics}.  Thus, being impossible to know $\tau$, the   mean square error (MSE), $\sigma^{2}_{\mathrm{MPI}}  \overset{\Delta}{=} \mathrm{E } (y-\hat{b})^2$ , of the "most" ideal MPI mitigation is  
 \begin{equation}\label{eq:mpiVar}
\sigma^{2}_{\mathrm{MPI}}  = \sum_{X} \rho^2 V_{b}^2\overline{B^2(t)}(\sqrt{1+X/V_{b}}-1)^2   + \sigma^{2}_{\mathrm{o}} 
\end{equation}
 where $X=\{-3,-1, 1, 3\}$, $\sigma^{2}_{\mathrm{o}} $ is the AWGN noise variance. The analytical form of BER with pure MPI and AWGN can be modified as 
\begin{equation}\label{eq:eq1a}
    \mathrm{BER}=\frac{2(M-1)}{M\mathrm{log}_2 M}Q\left(\sqrt{\frac{6\mathrm{log}_2 M}{M^2-1} \gamma}\right)
\end{equation}
    % \mathrm{BER}=\frac{1}{\sqrt{2\mathrm{\pi}}}\int\limits_{Q}^{+\infty}\exp{\left(-\frac{x^2}{2}\right)}\,\mathrm{d}x
where 
\begin{equation}\label{eq:eq1}
    \gamma = \frac{    \left(\mathrm{SNR}_{\mathrm{o}}^{-1} + \sigma_{\mathrm{MPI}}^2/ \mathrm{E }|d|^2 \right)^{-1}  }{2 \mathrm{log}_2 M} 
\end{equation}
where $M = 4$, $\mathrm{SNR}_{\mathrm{o}}$ is the SNR contributed by AWGN only with respectively to the signal power  (without MPI).
\begin{figure}[t!]
    \centering
    \includegraphics[height=6cm]{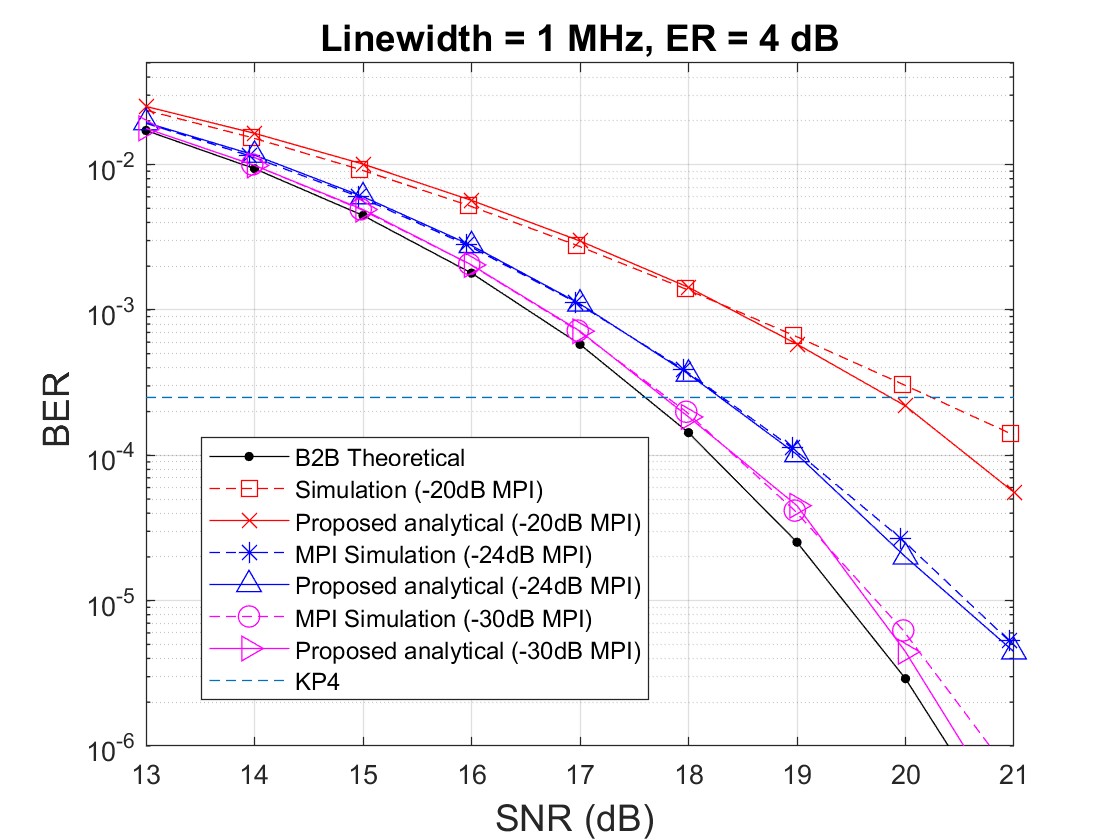}
    \caption{Comparison between simulation and the proposed analytical BER. 1 MHz Laser linewidth, ER = 4dB.}
    \label{fig:BERvsSNR}
\end{figure}

\begin{figure}[t!]
    \centering
    \includegraphics[height=6cm]{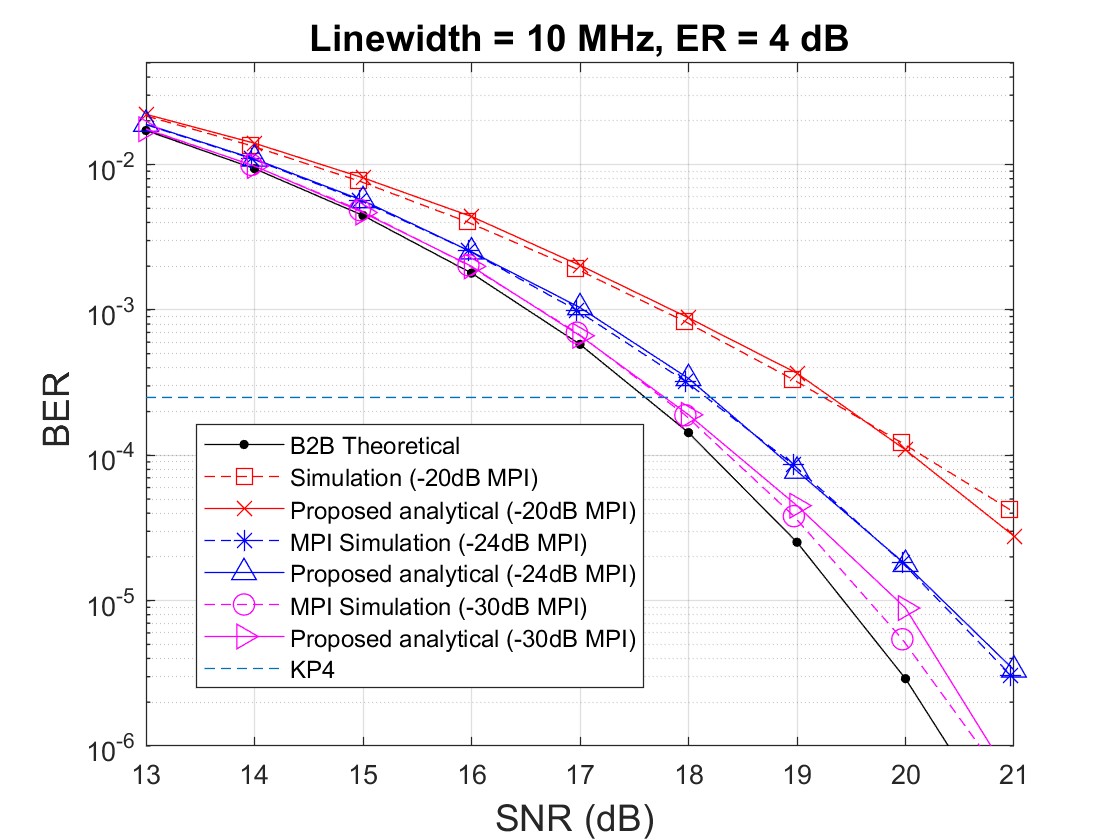}
    \caption{Comparison between simulation and the proposed analytical BER. 10 MHz Laser linewidth, ER = 4dB.}
    \label{fig:BERvsSNR2}
\end{figure}
 \section{Numerical Verification}
To verify our proposed achievable BER under MPI, 106.25GBaud gray-coded PAM4 data was generated digitally and Wiener laser phase noise $e^{j \theta}$ can be multiplied in the field domain (squar root of amplitudes). To generate reflection, a replica of the signal field was attenuated by a specified MPI ratio, $\rho^2$, and delayed by 2 km. As the laser linewidth is in MHz range, the laser phase noises of signal and reflection paths are effectively de-correlated and can be generated independently without applying delay. Assume zero fiber birefringence and the wavelength of the optical signal is at zero dispersion wavelength. To account for only the MPI impairment, ISI-free devices were assumed. At the receiver side, AWGN was added with respect to the signal power only (note that MPI should not be included in the signal power during AWGN generation). The two paths were added at photodetector via a square law device. At one sample per symbol, assuming perfect timing phase recovery, the mean of received PAM4 signal was removed, followed by renormalization. MPI was mitigated using the best bias estimate in Eq.~(\ref{eq:biasEst}) to examine the achievable performance in the presence of MPI. Lastly, decoding and bit error rate calculation were performed. 
\par 
Fig.\ref{fig:BERvsSNR} shows the comparison between our proposed analytical form and the numerical results, for various MPI ratios and linewidth of 1MHz. The same comparison for 10MHz linewidth is shown in Fig.\ref{fig:BERvsSNR2}. One can observe that both simulation and our proposed analytical form agree with each other at BER higher than $10^{-4}$. The discrepany appeared at BER below than $10^{-4}$ is caused by the insufficient number of bits (only 1 million bits were used). 
\begin{figure*}[t]
    % \centering
    \includegraphics[height=4.8cm]{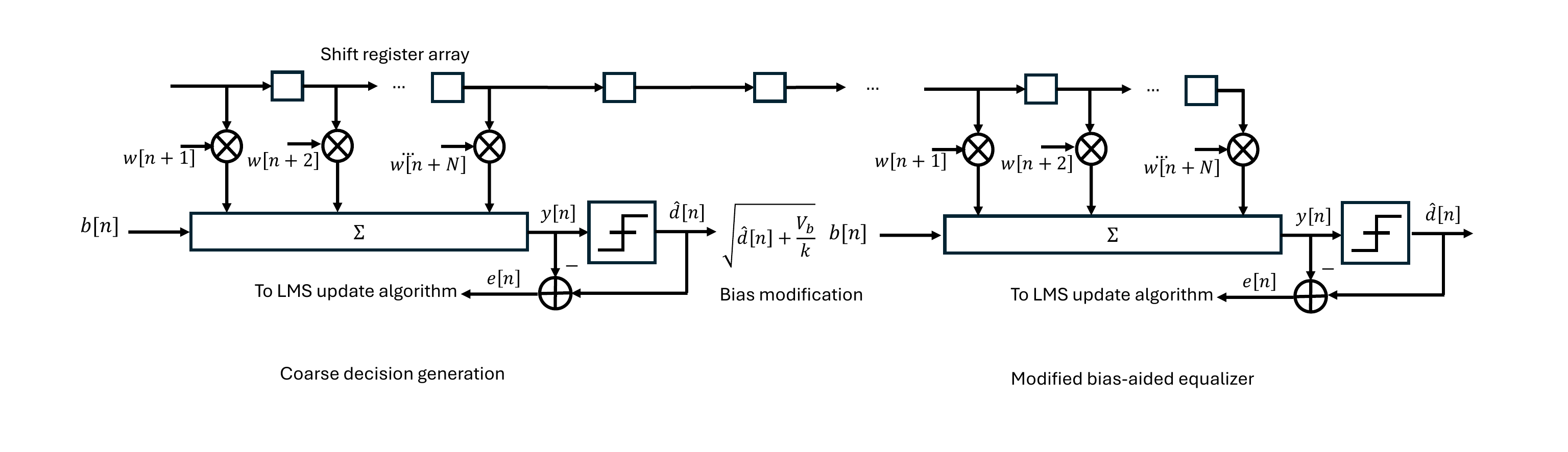}
    \caption{Joint ISI and MPI mitigation}
    \label{fig:dsp}
\end{figure*}
\section{Insight from Ideal Bias Estimate}
In the previous section, the ideal bias fluctuation estimate would bring us insight on how the current FFE can be modified to achieve better results. First, the error $e(t) = y(t) - d(t)$, required by an equalizer, using Eq.~(\ref{eq:rxSignal}), is

\begin{equation}\label{eq:errTerm}
\begin{split}
e(t) = 2 \rho \sqrt{ V_{b} + d(t)} \cdot \sqrt{V_{b} +d(t-\tau)} B(t) + n(t).
\end{split}
\end{equation}
The slowly varying  envelope, $B(t)$, can be tracked from the error averaged over the time scale of Baudrate to remove the Baud-rate interference $d(t-\tau)$, if  $d(t)$ and $V_b$ are known:   
\begin{equation}\label{eq:eq3}
\overline{e(t)} \cong 2 \rho \sqrt{d(t) + V_b}    \sqrt{V_b} B(t).   
\end{equation} 
This requires either a adata-assisted approach, or one must jointly estimate both $d(t)$ and the fluctuation $B(t)$. 
To this end, Fig. \ref{fig:dsp} suggests an architecture consisting of a coarse decision generation and a modified bias-aided equalizer. As a decision-directed approach relies on correct bipolar decisions $\hat{d}(t) = \{-3, -1, 1, 3\}$, the first stage in Fig. \ref{fig:dsp} is an ordinary FFE with bias adaptation. However, only the FFE decisions will be used by the subsequent equalizer, with a suitable rescaling based on the decision and $V_b$. The DD-LMS equations for the second-stage equalizer are thus 
\begin{equation}\label{eq:LMS}
\vec{w}[n+1] = \vec{w}[n] - \mu_w e[n] \vec{y}[n]   \in \mathbb{R}^N
\end{equation}
\begin{equation}\label{eq:eq5}
b[n+1] = b[n] - \mu_b e[n]  \sqrt{ \hat{d}[n] + V_b }     \in \mathbb{R}
\end{equation}
where $\vec{w}[n] = [ w[n+1],  ..., w[n+N]]^T$, are the taps for compensating ISI with step size $\mu_w$, while $\vec{y}[n] = [ y[n+1], ..., y[n+N]]^T$ is the received ADC signal. $2 \rho \sqrt{V_b}$, being a constant in Eq. (\ref{eq:eq3}), is absorbed into bias step size $\mu_b$. Note that there should be well-designed shift registers to align the decisions from the first-stage FFE with the received ADC signals entering the second-stage FFE.

% \section{Citing prior work}
% References appear in the text as superscripts like this\cite{example:article0}, this This saves some space, which can come in handy on a 3-page paper.
% this\cite{example:OFC2025},

\begin{figure}[t!]
    \centering
    \includegraphics[height=3cm]{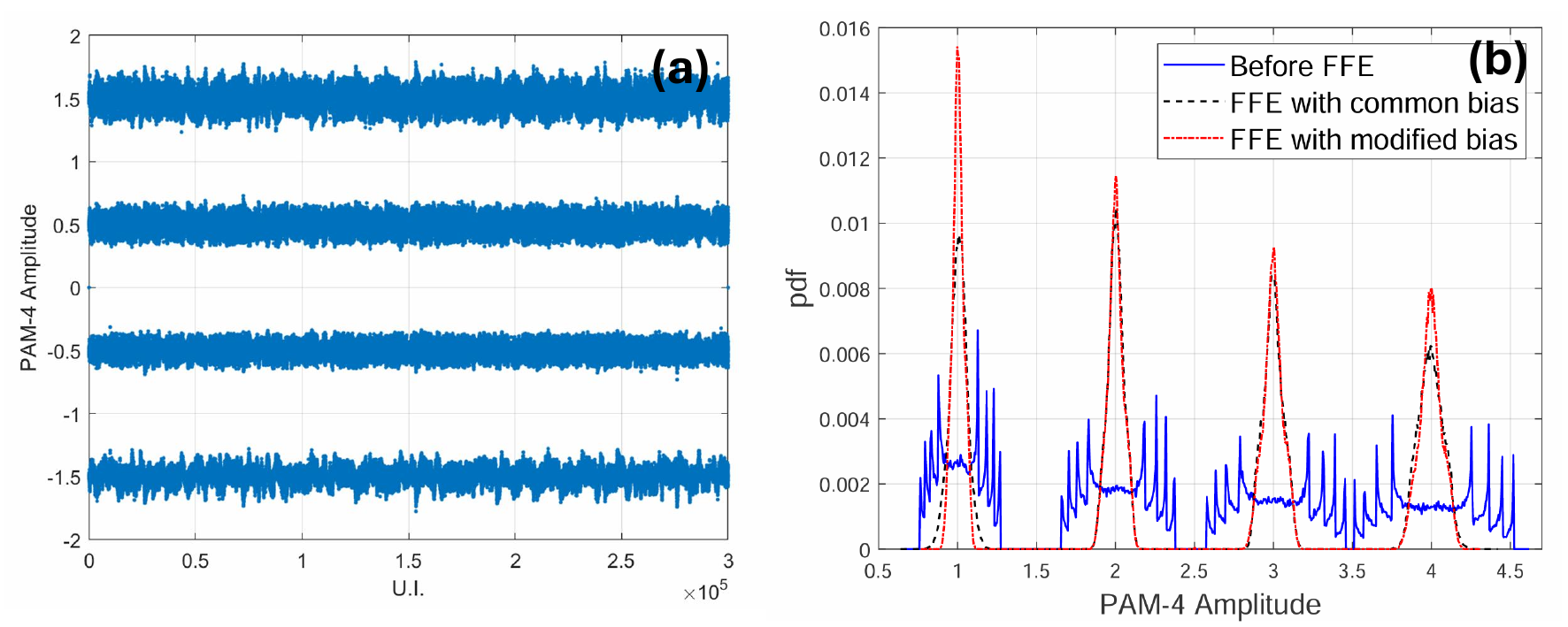}
    \caption{ (a) FFE with common bias compensation (b) Comparison mong distributions of waveforms before FFE, after FFE with common bias compensation, and after FFE with modified bias compensation}.
    \label{fig:dspResults}
\end{figure}

Fig. \ref{fig:dspResults} shows the waveforms using the architecture in Fig. \ref{fig:dsp} for a noise-free, -24dB MPI channel. Fig. \ref{fig:dspResults}(a) shows the soft symbols from the front FFE in Fig. \ref{fig:dsp}. As the traditional approaches cancels the bias fluctuations for the four PAM4 levels equally (called "common bias compensation"), it transfers the MPI noise from the higher amplitude to the lower amplitude. The advantage of the second-stage FFE with bias adjustment in Eq. (\ref{eq:LMS}) can be visualized by the soft symbol distribution in Fig. \ref{fig:dspResults}(b), where the post FFE shrinks both the lowest- and higest-level distributions via the bias scaling in Eq.~(\ref{eq:LMS}).

%-------------------------------------------------- Figures Section  -------------------------------------------------------%

%  \begin{figure}[t!]
%     \centering
%     \includegraphics[height=3cm]{waveformDistr.pdf}
%     \caption{ (a) FFE with common bias compensation (b) Comparison mong distributions of waveforms before FFE, after FFE with common bias compensation, and after FFE with modified bias compensation} (Noiseless, MPI = -24dB).
%     \label{fig:oldDSP}
% \end{figure}

\section{Conclusions}
In this work, we explore the achievable performance of an MPI channel for a single-reflection case. By "achievable" we refer to the best BER that can be achieved by the linear equalizer (FFE) in the standard DSP architecture. Based on our mathematical formulation, we construct the "ideal" estimate of the beat term (that causes bias fluctuation) to cancel the MPI. This leads us to obtain the residual MPI variance to calculate the BER analytically.  Finally, we discuss that this "ideal" bias estimate brings us insight on how to further optimize the current DSP architecture for PAM4. 

%-------------------------------------------------- Acknowledgements Section -------------------------------------------------------%
\clearpage

%-------------------------------------------------- Bibliography Section -------------------------------------------------------%
% see also https://tex.stackexchange.com/questions/55030/text-before-references-but-after-bibliography-title-with-bibtex as of 2024-02-29
\defbibnote{myprenote}{}

\printbibliography[prenote=myprenote]

\vspace{-4mm}
% \bibliography{references}
%\bibliography{C:/Users/autum/Desktop/ECOC2025/ECOC_2025_LaTeX_Template/references.bib}
%%%%%%%%%%%%%%%%%%%%%%%%%%%%%%%%%%%%%%%%%%%%%
%---------------------------------------------- End of Document -----------------------------------------------%
\end{document}